\documentclass[12pt,a4]{article}
\usepackage{epsf,epsfig,rotating}
\usepackage{latexsym,amssymb,amsfonts}
\begin{document}

\begin{center}

{\bf \large $\mathbf {K^0 - \bar K^0, \; B^0 - \bar B^0}$ mixings in the MSSM \\ 
with explicit $\mathbf {CP}$ violation in the Higgs sector}

\vspace{10mm}

M.N. Dubinin\\
{\it Institute of Nuclear Physics, Moscow State University, 119899 Moscow, Russia}
\footnote{ \hspace*{3mm} email: dubinin@theory.sinp.msu.ru}\\
\vskip 3mm
A.I. Sukachev\\
{\it Physics Department, Moscow State University, 119899 Moscow, Russia}
\footnote{ \hspace*{3mm} email: salex-82@yandex.ru }\\

\vspace{15mm}

\end{center}

\begin{quote}

{\footnotesize We consider the $K^{0}$ --- ${\hat K^{0}}${} and $B^{0}$ --- ${\hat B^{0}}${} mixings in the MSSM with the two-Higgs-doublet scalar sector featuring
explicit CP violation, and the Yukawa sector of type II.
In the case of strong mixing between CP-odd and CP-even states the existence of
light charged Higgs is allowed in the model. The mass splitting $\Delta m_{LS}${} and the amount of
indirect CP violation $\varepsilon${} are calculated.  In the limit of effective low-energy
approximation the nonstandard effects are shown to be negligibly small
in $\Delta m_{LS}${} and $\varepsilon${} for the $K^0$-mesons, being almost independent on the
charged Higgs mass. However, for the $B_{d}^{0}$---$\bar {B_{d}^{0}}$ and $B_{s}^{0}$---$\bar
{B_{s}^{0}}${} systems the effects of nonstandard physics are shown to be larger,
limiting the MSSM parameter space.}

\end{quote}

\begin{center}
{\bf PACS: 12.60.Jv; 14.80.Cp}
\end {center}

\vspace{15mm}

\section{Introduction}

\hspace*{6mm} Minimal supersymmetric extension (MSSM) of the standard model (SM) \cite{nilles}
involves many sources of $CP$ violation besides the well-known CKM mixing.
In the general case a number of complex parameters can be included in the MSSM leading to
the observable effects of CP violation dependent on their phases \cite{susyphases,cernrep} 
in addition to the only CKM phase of the SM. 

The number of phases in the soft SUSY breaking terms can be reduced significantly using the assumption of trilinear parameters
$A_{t,b}$ universality in the Higgs boson-squark sector and neglecting the gaugino phases.
Restricted $CP$ violation \cite{dub} of the Higgs potential can be described by the only one universal
phase  $\varphi = {\tt arg} (\mu A_{b}) = {\tt arg} (\mu A_{t})${} in addition to the CKM phase
($\mu$ is the Higgs superfield mixing parameter) and leads to a relatively simple picture of  radatively induced $CP$ violation in the MSSM two-doublet Higgs potential. A calculation of radiative corrections by means of the effective potential method leads to the complex parameters
$\lambda_{5,6,7}$ depending on the products $\mu A_{t,b}$ with phases different by a factor of two,
${\tt arg}\lambda_5=2\, {\tt arg} \lambda_{6,7}$. Direct experimental reconstruction of phases could be performed in the processes of superpartners and Higgs bosons production at the colliders 
\cite{sparticles} while an indirect evidence could be given by measurements of electric dipole moments \cite{edm}, mixings of the neutral mesons and meson decays \cite{mesons}.

The MSSM two-doublet Higgs sector contains three neutral scalars and two charged. Radiatively induced complex parameters of the Higgs potential, when appear, lead to explicit $CP$ violation and the three neutral scalars  $h_1, \; h_2, \; h_3${} which are mixings \cite{add, general} of $CP$-even states $h$, $H$ and $CP$-odd state $A$ known in the limit of $CP$ conserving two-doublet potential. The lightest scalar $h_1$ can have a mass substantially smaller than the LEP2 direct limit for the SM  $m_H >$114 GeV \cite{smlimit} being not observed at LEP2 energies because of the $ZZh_1${} coupling suppression 
\cite{higgs1} by parameters of the $h_1, \; h_2, \; h_3${} mixing matrix  which does not require specifically large values of the ${\tt tg} \beta$=$v_2/v_1$ parameter. Charged scalars $H^\pm$ could be relatively light, with masses 50-100 GeV. In the framework of such MSSM scenario an additional box diagrams with charged Higgs exchanges could contribute noticeably to mass splittings and mixing parameters in the neutral meson systems.
In this connection it is interesting to analyse MSSM implications to the neutral meson mixings, the only systems where effects of $CP$ violation are experimentally observed.
 
In the following first we consider the neutral $K^0$-meson system where $CP$ violation was discovered  
\cite{fitch}. The approximations used in this study are also applied to the $B_{S}^{0}, \; B_{D}^{0}$ mesons. In section II we briefly describe the MSSM Higgs sector and the Yukawa sector of type II (model MSSM II) and approximations used in the SM for calculation of the meson mass splittings $\Delta m_{LS}${} and the $CP$ violation parameter $\varepsilon${} of the neutral $K$-meson system. In sections 3,4,5 we calculate $\Delta m_{LS}${} and $\varepsilon${} including the one-loop box diagrams with one or two charged Higgs boson exchanges using the low-energy approximation. Some consequences are discussed in section 6.    

\vspace{5mm}

\section{Model MSSM II}

\hspace*{6mm}{\it \bf Yukawa sector.}
The Yukawa sector Lagrangian in a model with $n$-doublet ($\Phi_{n}$) structure of the scalar sector can be written in the form
\begin{eqnarray}
\Lambda_{Y} = - \sum_{ijn} g^{u \; n}_{ij} \overline{\left( \begin{array}{c}
u'_{iL} \\
d'_{iL}
\end{array} \right)} \left( \begin{array}{c}
\phi_{n}^{+} \\
\phi_{n}^{0}
\end{array} \right) u'_{jR} - \sum_{ijn} g^{d \; n}_{ij} \overline{\left( \begin{array}{c}
u'_{iL} \\
d'_{iL}
\end{array} \right)} \left( \begin{array}{c}
\phi_{n}^{0*} \\
-\phi_{n}^{+*} 
\end{array} \right) d'_{jR}
\label{eq9} 
\end{eqnarray}
plus hermitian conjugated term. Here $g^{u \; n}_{ij}, \; g^{d \;
n}_{ij}$ --- are the Yukawa couplings of up and down quarks with scalar doublets
which are $3 \times 3$-matrices in the flavor space with matrix elements wich are, generally speaking, complex, $\{u'_{iL}, \; d'_{iL}\}$ are the left doublets of up and down quarks, $u'_{jR}$ are right singlets of up quarks and $d'_{jR}$ are right singlets of down quarks ($i, j=${}1,2,3 define the quark generation number).

The two-doublet Yukawa sector of type II \cite{weinberg-0} does not include the 
$g_{ij}^{u \; 2}, \; g_{ij}^{d \; 1}$ terms, so $\Phi_{1}${} doublet generates only the up quark masses and $\Phi_{2}${} doublet only the down quark masses. The MSSM II Yukawa sector Lagrangian 
is given by
\begin{equation}
-L_{Y}^{II} \; = \; g^{u \; 1}_{ij} {\bar Q'}_{i \; L} {\tilde \Phi_{1}} u'_{j \; R} \; + \; g^{d \;
2}_{ij} {\bar Q'}_{i \; L} \Phi_{2} d'_{j \; R} \; + \; \mathrm{lept}. \; \mathrm{sec}. \; + \;
\mathrm {h.c.}
\end{equation} 

In terms of physical fields (diagonalized Lagrangian) the neutral scalars $A, \; H, \; h${} participate only in the flavor-conserving quark interactions, the charged scalars take part only in the flavor-changing transitions. The flavor-changing neutral currents do not appear. 

{\it \bf Scalar sector.}
The general hermitian renormalizable $SU(2)$ $\otimes U(1)$-invariant MSSM potential
at the $m_{\mathrm{top}}$ energy scale has the form
\begin{equation}
U(\Phi_1,\Phi_2) = - \, \mu_1^2 (\Phi_1^\dagger\Phi_1) - \, \mu_2^2 (\Phi_2^\dagger \Phi_2) -
\mu_{12}^2 (\Phi_1^\dagger \Phi_2) - \stackrel{*}{\mu_{12}^2} (\Phi_2^\dagger \Phi_1) +
\label{eq:genU}
\end{equation}
$$ 
+ \lambda_1 (\Phi_1^\dagger \Phi_1)^2 + \lambda_2 (\Phi_2^\dagger \Phi_2)^2 + \lambda_3
(\Phi_1^\dagger \Phi_1)(\Phi_2^\dagger \Phi_2) + \lambda_4 (\Phi_1^\dagger \Phi_2)(\Phi_2^\dagger
\Phi_1) + 
$$
$$ 
+ \frac{\lambda_5}{2}(\Phi_1^\dagger \Phi_2)(\Phi_1^\dagger\Phi_2)+\frac{\stackrel{*}{\lambda}_5}
{2}(\Phi_2^\dagger \Phi_1)(\Phi_2^\dagger \Phi_1) + 
$$
$$
+ \lambda_6 (\Phi^\dagger_1 \Phi_1)(\Phi^\dagger_1 \Phi_2)+ \stackrel{*}{\lambda}_6(\Phi^\dagger_1
\Phi_1)(\Phi^\dagger_2 \Phi_1) + \lambda_7 (\Phi^\dagger_2 \Phi_2)(\Phi^\dagger_1
\Phi_2)+\stackrel{*}{\lambda}_7(\Phi^\dagger_2 \Phi_2)(\Phi^\dagger_2 \Phi_1).  
$$

At the $M_{SUSY}${} mass scale (energy of the order of sparticle masses)
$\lambda_{1, \dots, 7}${} parameters are real and expressed through the electroweak $SU(2) \otimes U(1)$ couplings $g_{1}${} and $g_{2}${} \cite{inoue}. At the energy scale $m_{\mathrm{top}}$ large radiative corrections to the scalar self-interactions can appear from the box and triangle one-loop diagrams with scalar quark exchanges, generating complex $\lambda_{5, \dots, 7}$ parameters. In the restricted complex MSSM mentioned in the Introduction, these corrections depend on four parameters, the trilinear universal $A_{t,b}$, Higgs superfield parameter $\mu$, SUSY mass scale $M_{SUSY}${} and the universal phase $\varphi={\tt arg} (\mu A_{t,b})$. 

The diagonalization procedure for the general two-doublet potential, leading to Higgs boson mass eigenstates and their self-interactions in the physical basis, can be found in \cite{add, dubinin-1}. 
Three of the eight scalar states from two complex doublets (the Goldstone modes $G^{0}${} and
$G^{\pm}${}) are eaten by $W^{\pm}$-{} и $Z^{0}$-{} bosons, giving two charged scalars
$H^{\pm}${} and three neutral scalars $h_{1}, \; h_{2}, \; h_{3}$ without definite
$CP$-parity which are the mixtures of $h, \; H, \; A$. The charged Higgs boson mass is given by
\cite{dub,dubinin-1}
\begin{equation}
m^{2}_{H^{\pm}} \; = \; m_{W}^{2} + m_{A}^{2} - \frac{v^{2}}{2}( 
{{\tt Re} \Delta \lambda_{5}} - \Delta
 {\lambda_{4}}),
\end{equation}
where $m_W$ is the $W$ boson mass, $m_A$ is the $CP$-odd scalar mass defined at the zero phase, $v=2\,
m_W/g_2$, and the effective parameters $\Delta \lambda_{4,5}$ at the one-loop are \cite{dub}
\begin{eqnarray}
\label{eq:lambda4}
\Delta \lambda_4 &=& - \, \frac{3\, g_2^2}{32\pi^2}\,(h_t^2+h_b^2)\, \,\ln\left(
\frac{ M^{\,2}_{\rm SUSY}}{m_{\mathrm{top}}^2}\right) + \frac{3}{8\pi^2}\,h_t^2h_b^2\left[\
\ln\left(\frac{ M^{\,2}_{\rm SUSY}}{m_{\mathrm{top}}^2}\right) + \frac{1}{2} X_{tb}\right]\, 
 \\ \nonumber 
&&  -\, \frac{3}{96\pi^2}\, \frac{|\mu|^2}{M^2_{\rm SUSY}} \left[ h^4_t\,  \left(\, 3\, -\, \frac{
A_t|^2}{M^2_{\rm SUSY}}\, \right)\, +\,  h^4_b\, \left(\, 3\, -\, \frac{|A_b|^2}{M^2_{\rm SUSY}}\,
\right)\right] \,  \\ \nonumber
&& + \,\, \frac{3 g_2^2\left[h_b^2(|\mu|^2-|A_b|^2)+h_t^2(|\mu|^2-|A_t|^2)\right]}{64\pi^2
M^{\,2}_{\rm SUSY}} + \, \frac{3 g_2^4}{64 \pi^2} \,\,\ln\left(\frac{ M^{\,2}_{\rm SUSY}}
{m_{\mathrm{top}}^2}\right) \,, 
\end{eqnarray}
\begin{eqnarray}
\label{eq:lambda5}
\Delta\lambda_{\,5} &=&  \, \frac{3}{96\,\pi^2}\, \left(h^4_t\, \left(\frac{\mu A_t}{M^{\,2}_{\rm
SUSY}}\right)^2\, +\, h^4_b\, \left(\frac{\mu A_b}{M^{\,2}_{\rm SUSY}}\right)^2\right) 
\end{eqnarray}
The Yukawa couplings $h_{\,t} = \frac{\sqrt{2}\, m_{\,t}}{v \sin\beta }$, $h_{\,b} =
\frac{\sqrt{2}\, m_b }{v \cos\beta }$ and the $X_{tb}$ stands for
\begin{equation}
X_{tb}\equiv\frac{|A_t|^2+|A_b|^2+2{\tt Re}(A_b^*A_t)}{2M^{\,2}_{\rm SUSY}}-\,\frac{|\mu|^2}
{M^{\,2}_{\rm SUSY}}-\,\frac{||\mu|^2-A_b^*A_t|^2}{6M^{\,4}_{\rm SUSY}}\, .
\end{equation}

Radiative corrections to the effective potential defined by (\ref{eq:lambda4}) and (\ref{eq:lambda5}) at the scale $m_{\mathrm{top}}$
\footnote{It is essential that (5) and (6) are valid for insignificantly different squark masses \cite{close}. The wave-function renormalization terms in $\Delta \lambda_4$ \cite{dub,add} are omitted. For the case of large difference between the squark masses another decomposition of the effective potential must be used, see \cite{diffsq}.}
are especially large at moderate $M_{SUSY}$ (several hundreds of GeV) and large $\mu$, $A_{t,b}$ parameters (of the order of 1 TeV). The scalar quark sector is strongly coupled. For the case of substantial phase ${\tt arg}(\mu A_{t,b})$ the $h_{1,2,3}$ mass splittings at moderate $H^\pm$ mass of 150-180 GeV are as large as 10-50 GeV. Unacceptable regions of the MSSM parameter space exist, where $m_{h_{1,2,3}}${} are close to zero or not positively defined. Note that the correction (\ref{eq:lambda4}) includes additional terms in comparison with the useful approximation (see, for example, \cite{boos}) $m_{H^\pm}^2= m^2_A + m^2_W
-\epsilon$, where $\epsilon = 3\, G_F m^2_W/4 \pi^2 \sqrt{2}\, (m^2_{t}/\sin^2 \beta + m^2_b/\cos^2
\beta) \ln (M^2_{SUSY}/m^2_{\mathrm{top}})$ corresponds to the first term only in the expression for
$\Delta \lambda_4$, see (\ref{eq:lambda4}).

Charged Higgs boson masses in the framework of typical MSSM scenario with strong mixing of the neutral
$CP$-even/$CP$-odd states (known as the CPX scenario, see \cite{higgs1}) are shown in Fig.1. The two-loop corrections to $\lambda_4$ and the nonleading D-terms \cite{dub} are accounted for.
In the CPX scenario under consideration the constraint on $M_{SUSY}$, $A_{t,b}$ and $\mu$ is imposed as
$\mu=2\, A_{t,b}= 4\, M_{SUSY}$, everywhere in the following $M_{SUSY}=$ 500 GeV. At moderate phases ${\tt arg}(\mu A_{t,b})$ and small ${\tt tg}\beta$ the lightest neutral scalar mass is positively defined, see Fig.1a, at relatively small $m_{H^\pm}$ starting from 20-30 GeV. The contours
at the lightest neutral scalar mass $m_{h_1}=40$ GeV, see Fig.1b, strengthen the  limit on $m_{H^\pm}$ by approximately 20 GeV. At the same time at moderate values of the phase ${\tt arg}(\mu A_{t,b})$ of the order of 10-15 degrees the lightest neutral Higgs boson $m_{h_1} \sim 40$--$50${} GeV becomes poorly observable at LEP2 luminosities in the channel  $e^+ e^- \to Z h_1$ owing to the increase of the $A$ scalar state $CP$-odd admixture in the $h_1$ state with the suppression of the $ZZh_1$ coupling. The situation depends critically on the $m_{\mathrm{top}}$ value, at the top quark mass greater than 178 GeV and moderate ${\tt tg}\beta \sim$3--10 any value of $m_{h_1}$ could be possible down to values close to zero \cite{bechtle}. Small deviations from the CPX scenario with the $A_{t,b}$ and $\mu$ parameters shifted by 100 GeV are illustrated in Fig.2a and Fig.2b. In this case the region of smallest possible charged scalar masses is displaced to the large  ${\tt tg}\beta \sim${} 40---45 interval. We are not considering in the following an extremely large ${\tt tg}\beta$ values exceeding 50 insofar as such values become to be poorly consistent with the Tevatron limits \cite{tevatron} on the $t\to H^+ b$ decay channel. The $t\to H^+ b$ decay probability is enhanced by ${\tt tg}\beta$ factor in the vertex and becomes poorly consistent with direct experimental data on the top quark pair production with the following decay $t\to W^+ b$. Direct LEP2 experimental data on the $m_{H^\pm}$ production in the channels $e^+ e^- \to c \bar s \bar c s$, $c \bar s \tau^- \bar \nu_{\tau}$ $\tau^+ \nu_{\tau} \tau^- \bar \nu_{\tau}$ gives the limit $m_{H^\pm}>$56 GeV \cite{delphi}. Note that the signal cross section is sensitive to potentially large one-loop corrections coming from superpartners and gluino \cite{kraml}. Specific event charasteristics in the channel $H^\pm \to W^\pm h_1$ for a light charged Higgs produced in the top decay at the LHC and discovery possibilities were analysed in \cite{godbole}.

\section{ $K^0$-$\bar K^0$ mixing in the SM}

{\it \bf $K^0$-$\bar K^0$ mixing in the SM.} The neutral kaon mixing in vacuum is characterized by the mass splitting of physical states $\Delta m_{LS}${} and by the parameter of $CP$ violation $\varepsilon${}. In the framework of phenomenological approach to the mixing, see, for example, 
\cite{vysotsky1}, the effect is described by the nonhermitian Hamiltonian (mass matrix) of dimension 2$\times$2 and of the form $M_{ij}-\frac{i}{2}\Gamma_{ij}$, where the real $M_{12}$ and the imaginary $\Gamma_{12}$ parts of nondiagonal matrix elements define the value of $\Delta m_{LS}$ and the weights of
$CP$-even/$CP$-odd states in the physical states of neutral mesons. The amount of $CP$ violation in the phenomenological approach to the meson mixing is given by a small deviation from the ortogonality of the eigenvectors in the physical basis (given by meson states without definite $CP$-parity which appear after the diagonalization of 2$\times$2 Hamiltonian). The real and the imaginary parts of the
off-diagonal matrix elements are defined by box diagrams (see Fig.3) with the intermediate $u$, $c$, $t$ quarks, and box diagram with the cut ($c$ and $t$ excluded), respectively.
The SM mixing in the neutral kaon system appears as a consequence of mixing in the charged current sector defined by the CKM matrix \cite{maskawa}, and a small value of the mixing is explained by the GIM mechanism  \cite{glashow}, see Fig. 3.

The real part of the amplitude defines the mass splitting 

$$
\Delta m_{LS}^{WW} \; = \; \frac{G^{2}_{F}f_{K}^{2}m_{K}B_{K}}{6{\pi}^{2}} \mathrm{Re}
\Biggl[ (V_{cd}^{*})^{2}V_{cs}^{2}m_{c}^{2}\eta_{1} I({\xi_{1}}) \; + 
$$
\begin{equation}
\left. + \; (V_{td}^{*})^{2}V_{ts}^{2}m_{t}^{2}\eta_{2} I({\xi_2}) \; + \;
2V^{*}_{td}V_{cd}^{*}V_{ts}V_{cs}\eta_{3} \frac{m_{c}^{2}m_{t}^{2}}{m^{2}_{t} - m^{2}_{c}} \;
\mathrm{ln} \frac{m^{2}_{t}}{m_{c}^{2}} I(\xi_{1}, \xi_{2}, \xi_{3}) \right].
\label{eq12}
\end{equation}
The imaginary and the real parts ratio defines indirect $CP$ violation parameter
\begin{equation}
|\varepsilon| \; = \; \frac{1}{2\sqrt{2}} \frac {Im \; A}{Re \; A}, 
\end{equation}
\begin{eqnarray}
A \; = \; \left[ (V_{cd}^{*}V_{cs})^{2}m_{c}^{2}\eta_{1} I(\xi_{1}) \; + \;
(V_{td}^{*}V_{ts})^{2}m_{t}^{2}\eta_{2}I(\xi_{2}) \right. \; + \nonumber \\ 
+ \; \left.  2 \; V^{*}_{td}V_{cd}^{*}V_{ts}V_{cs}\eta_{3} \frac{m_{c}^{2}m_{t}^{2}}{m^{2}_{t} -
m^{2}_{c}} \; \mathrm{ln} \frac{m^{2}_{t}}{m_{c}^{2}} I(\xi_{1}, \xi_{2}, \xi_{3}) \right], 
\label{eq13}
\end{eqnarray}
where $\xi_{1} = (\frac{m_{c}}{m_{W}})^{2}, \, \xi_{2} = (\frac{m_{t}}{m_{W}})^{2}, \, \xi_{3} =
(\frac{m_{t}}{m_{c}})^{2}$  fix the value of Vysotsky \cite{vysocki2} and Inami-Lim \cite{inami} function $I(\xi)$
\begin{equation}
I(\xi) \; = \; \left \{ \frac{{\xi}^{2} - 11\xi + 4}{4(\xi - 1)^{2}} - \frac{3{\xi}^{2}\; ln \xi}{
(1-\xi)^{3}} \right \}, 
\label{in-1}
\end{equation}
taking into account the contributions of the order of $(m_t/m_W)^2$ and $(m_c/m_W)^2$ (the latter are small, so $I(\xi_{1}) \, \approx \, 1$),
and $I(\xi_{1}, \xi_{2}, \xi_{3})$ defines the contribution of "mixed" $ct$ exchange diagrams ($ct$-boxes) \cite{inami, urban}:
\begin{eqnarray}
I(\xi_{1}, \xi_{2}, \xi_{3}) \; \> = \> \; \left(\frac{\xi_{3}}{ln \, \xi_{3}} - \frac{1}{ln \;
\xi_{3}} \right) \left(\frac{ln \, \xi_{1}}{(1 - \xi_{1})^{2}(1 - \xi_{2})^{2}(1 - \xi_{3})} \; -
\; \right. \nonumber \\
- \; \frac{\xi_{1}}{(1 - \xi_{1})^{2}(1 - \xi_{2})^{2}} \; + \; \frac{(2 - \xi_{2})\xi_{2} \, ln \,
\xi_{1} - (2 - \xi_{1})\xi_{1} \, ln \, \xi_{2}}{(1 - \xi_{1})^{2}(1 - \xi_{2})^{2}(1 - \xi_{3})}
\; + \nonumber \\ 
\left. + \frac{\xi_{1}^{2}(1 - \xi_{2}) - \xi_{2}^{2}(1 - \xi_{1})}{(1 - \xi_{1})^{2}(1 -
\xi_{2})^{2}(1 - \xi_{3})} \> \right). \>
\label{in-2}
\end{eqnarray}
Here $f_{K} \, \approx 1.27\,  f_{\pi} \, \approx \, $165 MeV{} is the decay constant, $G_{F} \; = \; 1.17 \times 10^{-5} \;$ GeV$^{-2}$, $V_{ij}${} are the CKM matrix elements,
$B_{K} \; \approx \; 1.0${} is the nonperturbative QCD correction ("vaccuum insertion")
and $\eta_{1}, \; \eta_{2}, \; \eta_{3}${} are factorized perturbative QCD corrections \cite{vysocki2} from gluonic "cross-exchanges" connecting inner and outer quark lines (see also
\cite{vainshtein} for two generations case). They are expressed by means of effective QCD couplings ratio taken at the corresponding quark mass scale in the power defined by the anomalous dimension. 
The leading logarithmic (LL) corrections are sensitive to fermion masses and the energy scale where we replace the $W$-exchange amplitudes by the four-fermion ones, so corrections beyond LL are usually accounted for. For the kaon system most significant corrections are defined by $\eta_{1}$ because the main contribution is given by the $cc$ box diagram. In the following we are using the NLO QCD corrections for $\eta_{1}$, $\eta_{2}$,  $\eta_{3}$ calculated in \cite{herlich}. The NLO corrections to $\eta_{2}$ and $\eta_{3}$ contribute at the level of 10\% 
and are less important for the following consideration. We take $\eta_{1} = 1.3${} using the parametrization of \cite{herlich} at $m_c=$1.3 GeV and $\Lambda_{\bar {MS}}$=0.350 GeV, and $\eta_{2} = 0.47${}, $\eta_{3} = 0.57${}.
 
The contribution of the third generation quarks to $\varepsilon${} was obtained in \cite{vysocki2} and
\cite{inami}. The result (\ref{eq13}) includes both the contributions of the $cc$-box and the combined $ct$-box; (\ref{eq12}) accounts for the $tt$-box and the combined $ct$-box yield to the neutral kaon mass difference.

\section{$K^0$-$\bar K^0$ mixing in MSSM II.}

In the absence of flavor-changing neutral scalar exchanges, two additional box diagrams with one or two charged Higgs boson exchanges appear in the MSSM II, see Fig.\ref{fig4} ($HW$ diagram) and Fig.\ref{fig5} ($HH$ diagram). In the calculation of $HW$ exchange diagram using the t'Hooft-Feynman gauge, the unpysical scalar mode contribution was accounted for (Fig.7). Using the low-energy approximation $k^{2} \gg p^{2}_{i}${}, where $k$/$p_{i}$ is the inner/(outer leg $i$) momentum, the kaon mixing parameters can be evaluated as follows:   

\begin{equation}
\Delta m_{LS}^{HW} = \frac{G_{F}C_{H}f^{2}_{K}m_{K}B_{K}}{48 \pi^{2} m_{W}^{4}}\; \left( (m_{s} +
m_{d}) m_{H}^{2} \; \mathrm {Re} B_{1} + \frac{m_{H}^{2} m_{s}^{2}}{\sin 2\beta} \; \mathrm {Re}
B_{2} \right),
\label{eq19}
\end{equation}
\begin{equation}
\Delta m_{LS}^{HH} = \frac{f_{K}^{2}m_{K}B_{K}m^{2}_{s}}{48 \pi^{2}v^{4} m_{W}^{2}}\; \mathrm {Re}
\; C,
\label{eq18}
\end{equation}
где
$$
B_{1} \; = \; (V_{cd}^{*}V_{cs})^{2}m_{c}^{3}\eta_{4} I_{HW-1} (\xi_{1}, \xi_{4}, \xi_{6}) +
(V_{td}^{*}V_{ts})^{2}m_{t}^{3}\eta_{5} I_{HW-1} (\xi_{2}, \xi_{5}, \xi_{6}) \; +
$$ 
\begin{equation}
+ \; V_{cd}^{*}V_{td}^{*}V_{cs}V_{ts}m_{c}m_{t}(m_{c} + m_{t})\eta_{6} I_{HW-2} (\xi_{1}, \xi_{2},
\xi_{3}, \xi_{4}, \xi_{5}, \xi_{6}),  
\label{eq23}
\end{equation}
$$
B_{2} \; = \; (V_{cd}^{*}V_{cs})^{2}m_{c}^{2}\eta_{4} I_{HW-3} (\xi_{1}, \xi_{4}, \xi_{6}) +
(V_{td}^{*}V_{ts})^{2}m_{t}^{2}\eta_{5} I_{HW-3} (\xi_{2}, \xi_{5}, \xi_{6}) \; +
$$ 
\begin{equation}
+ \; 2 \cdot V_{cd}^{*}V_{td}^{*}V_{cs}V_{ts}m_{c}m_{t}\eta_{6} I_{HW-4} (\xi_{1}, \xi_{2}, \xi_{3},
\xi_{4}, \xi_{5}, \xi_{6}),  
\label{eq29}
\end{equation}
$$
C \; = \; (V_{cd}^{*}V_{cs})^{2}m_{c}^{2}\eta_{7} I_{HH-1} (\xi_{4}) +
(V_{td}^{*}V_{ts})^{2}m_{t}^{2}\eta_{8} I_{HH-1} (\xi_{5}) \; +
$$
\begin{equation}
+ \; 2 \cdot V_{cd}^{*}V_{td}^{*}V_{cs}V_{ts}m_{c}m_{t}\eta_{9} I_{HH-2} (\xi_{3}, \xi_{4},
\xi_{5}).
\label{eq22}
\end{equation}
The notation $C_{H}$ is used for the effective four-fermion coupling with scalar exchange, the
$G_{Fermi}${} analogue. Dimensionless functions $I_{HW-i}$ и $I_{HH-j}$,  $i = 1, \; 2,
\; 3, \;4; \; \; j = 1, \; 2$, are the analogues of Inami-Lim-Vysotsky functions, appearing
in the SM $W$-exchange diagrams, and $\xi_{i}$,
$i = 1,...,6$, are various mass ratios for particles of the internal lines,
$\xi_{1} = (\frac{m_{c}}{m_{W}})^{2}, \, \xi_{2} = (\frac{m_{t}}{m_{W}})^{2}, \,
\xi_{3} = (\frac{m_{t}}{m_{c}})^{2}, \, \xi_{4} = (\frac{m_{c}}{m_{H}})^{2}, \, \xi_{5} =
(\frac{m_{t}}{m_{H}})^{2}, \, \xi_{6} = (\frac{m_{H}}{m_{W}})^{2}$. Symbolic expressions for
$I_{HW-i}$ and $I_{HH-j}$ and more details can be found in the Appendix. We are using $\eta_{1} \, = \, \eta_{4} \, = \, \eta_{7} \, = \, 1.3${}, $\eta_{2} \, = \,
\eta_{5} \, = \, \eta_{8} \, = \, 0.47${} and $\eta_{3} \, = \, \eta_{6} \, = \, \eta_{9} \, = \,
0.57${}, defining the values of perturbative QCD corrections using \cite{herlich}.  
Numerical contribtions of $WW$, $HW$ and $HH$ amplitudes to the neutral kaons mass splitting and to the $\varepsilon${} parameter are shown in Table 1 and Table 2.
We define
\begin{equation}
|\varepsilon|_{LS}^{tot} \; = \; \frac{1}{2 \sqrt{2}} \frac{V_{LS}^{WW} \; + \; V_{LS}^{HW-1} \; +
\; V_{LS}^{HW-2} \; + \; V_{LS}^{HH}}{W_{LS}^{WW} \; + \; W_{LS}^{HW-1} \; + \; W_{LS}^{HW-2} \; +
\; W_{LS}^{HH}}
\label{eq41}
\end{equation}   
where $V_{LS}^{WW}, \; V_{LS}^{HW-1}, \; V_{LS}^{HW-2}, \; V_{LS}^{HH}${} и $W_{LS}^{WW}, \;
W_{LS}^{HW-1}, \; W_{LS}^{HW-2}, \; W_{LS}^{HH}${} are the imaginary and the real parts of various amplitudes multiplied by a corresponding effective factor
$$
V_{LS}^{WW} \; = \; G_{F}^{2} \cdot \mathrm{Im} \; A, \quad W_{LS}^{WW} \; = \; G_{F}^{2} \cdot
\mathrm{Re} \; A,
$$
$$
V_{LS}^{HW-1} \; = \; \frac{G_{F}C_{H}(m_{d} + m_{s})m_{H}^{2}}{8 m_{W}^{4}} \cdot \mathrm{Im} \;
B_{1}, \quad W_{LS}^{HW-1} \; = \; \frac{G_{F}C_{H}(m_{d} + m_{s})m_{H}^{2}}{8 m_{W}^{4}} \cdot
\mathrm{Re} \; B_{1},
$$
$$
V_{LS}^{HW-2} \; = \; \frac{m_{s}^{2}m_{H}^{2}G_{F}C_{H}}{8 m_{W}^{4} \sin 2\beta} \cdot \mathrm{Im}
\; B_{2}, \quad W_{LS}^{HW-2} \; = \; \frac{G_{F}C_{H}m_{s}^{2}m_{H}^{2}}{8 m_{W}^{4} \sin 2\beta}
\cdot \mathrm{Re} \; B_{2},
$$
$$
V_{LS}^{HH} \; = \; \frac{m_{s}^{2}}{v^{4}m_{W}^{2}} \cdot \mathrm{Im} \; C, \quad W_{LS}^{HH} \; =
\; \frac{m_{s}^{2}}{v^{4}m_{W}^{2}} \cdot \mathrm{Re} \; C.
$$

We have already mentioned the importance of $\eta_{1}$ QCD correction factor in front of the $cc$ box amplitude for the calculation of $\Delta m_{LS}${} и $\varepsilon${}. The NLO analysis of
\cite{herlich} gives the uncerainty of $\eta_{1}$ about 20\%: 
$\eta_{1} \, = \, 1.3 \pm 0.2$. We are using the central value $\eta_{1} \, = \, 1.3$. The values of
$\Delta m_{LS}${} and $\varepsilon${} for the SM contribution at different values of $\eta_{1}$ can be found in Table 3. Relatively small $HW$ and $HH$ amplitudes have small influence on the full contrubution with the variation of $\eta_{1}$ factor.

\section{$B_{d}^{0}$-$\bar B_{d}^{0}$ and $B_{s}^{0}$-$\bar B_{s}^{0}$ mixings in MSSM II}

There is no conceptual difference in the analysis of the neutral $B_{s}^{0}$- and $B_{d}^{0}$ meson systems in comparison with the analysis of the $K^0$ system. Mixing of the states without definite $CP$-parity in vacuum with the mass splitting and $CP$ violation defined by the CKM complex matrix elements and charged Higgs boson exchanges takes place. Numerical results can be obtained using formulae (\ref{eq12}), (\ref{eq13}), (\ref{eq19}), (\ref{eq18}), (\ref{eq23}), (\ref{eq29}) and (\ref{eq22}) given above with the replacement of $d$ or $s$ index by the $b$ index together with redefinition of the decay constant, "vacuum insertion" and meson mass. Different quark masses and mixing parameter values lead to qualitatively different numerical results. The yield of SM $cc$-box amplitude to $B^0$-meson mass splitting $\Delta m_{LS}^{B}$ is suppressed. Main contribution to mass splitting is given by the $tt$-box amplitude, the corresponding value of perturbative QCD correction factor $\eta_{B}$ is not critically different from the $\eta_{1}${} factor for the kaon system,  $\eta_{B} \; =\; 0.55$ \cite{buras}. The $u$ and $c$ quarks provide important contributions to the diagrams with a cut, defining the imaginary part of nondiagonal Hamiltonian terms $\Gamma_{12}$. Numerical values of $\Delta m_{B_{d}}$ и $\Delta m_{B_{s}}${} can be found in Table 4.  
One can observe that like for kaons, the contributions of $HW$ and $HH$ MSSM amplitudes to $B^0$ mass splittings are small. At $m_{H^\pm} = 50${} GeV the $HW$-box gives about 1\%
of $B^0$--$\bar B^0$ mass difference. The full splitting $\Delta m_{B_{s,d}} \; = \; \Delta m_{B_{s,d}}^{WW} + \Delta m_{B_{s,d}}^{HW} + \Delta m_{B_{s,d}}^{HH}${} is not significantly different from the SM result and is not in an excellent agreement with the experimental result. For example, in the SM we found $B^0_{d}$ mesons mass splitting $\Delta m_{B_{d}} \; = \; 2.12 \times 10^{-13}$ GeV  and the experimental result is $\Delta m_{B_{d}}^{exp} \; = \; (3.33 \pm 0.03) \times 10^{-13}$ GeV. Additional contributions of $HW$ and $HH$ box amplitudes, see Table 5, parametrically dependent on the charged Higgs boson mass, can be used to restrict it.

In order to evaluate the limits for $m_{H^\pm}$ we used the approach described in \cite{bityukov}.
Two statistical hypotheses are considered (1) charged Higgs boson exchanges contribute to the
mass splitting (2) these exchanges do not contribute. The value of statistical uncertainty $k$ is defined by {\it both}
probabilities (a) the probability to reject the hypothesis (1) when it is true, (b) the probability to accept (1) when (2) is true. The relative uncertainty is defined as $k=(a+b)/[2-(a+b)]$, see details in \cite{bityukov}, and converted to a number of standard deviations. For example, let us consider the case $m_{H^\pm}=$50 GeV for $B^0_d$-mesons (see Table 4). Then the SM mass splitting is 2.11$\times$10$^{-13}$ GeV and the negative contribution of charged Higgs exchanges is  -0.017$\times$10$^{-13}$ GeV. With the experimental accuracy 0.03$\times$10$^{-13}$ GeV indicated in the Table we get the relative uncertainty $k$ (i.e. the probability of wrong decision about the existence of charged Higgs yield) of 32$\%$ which corresponds to 0.5$\sigma$ level. If the experimental accuracy is taken to be 0.005$\times$10$^{-13}$ GeV which is six times better, the relative uncertainty becomes 2.5$\%$ corresponding to statistical significance of 2$\sigma$ (usually denominated as "the weak evidence" that the charged Higgs boson exchange contribution exists). With the experimental accuracy of the mass splitting measurement at the level of 0.003$\times$10$^{-13}$ GeV for $B^0_d$-mesons the restriction on the charged Higgs boson mass is $m_{H^\pm}>$112 GeV with the statistical significance of 2$\sigma$. The limit is stronger if the accuracy for $B^0_s$-mesons is improved by one order of magnitude to the level of 0.01$\times$10$^{-13}$ GeV (see Table 4) giving $m_{H^\pm}>$155 GeV at the 2$\sigma$ level. 

Note that for precise comparison of different amplitudes it is necessary to evaluate $W$,$H$-boson exchange diagrams at small distances. Our approximation $k^{2} \gg p^{2}_{i}$ is not precise enough (see also \cite{vysotsky1}) from this point of view leading to an underestimate in comparison with the experimental mass splitting. More exact estimates can be found in \cite{xiao, dias-2} where results greater by a factor of 1.2 -- 1.5 for the amplitudes have been obtained.   

The calculation of $B^0_{s,d}${} meson mixings is different from the case of $K^0$-mesons, because
the phase of leading term in the imaginary part $\Gamma_{12}${} is approximately the same as the phase of the leading term of the real part $M_{12}$, giving an additional suppression of the $\varepsilon${} parameter value by a factor $(m_{c} / m_{b})^{2}$. In this case the expression
(\ref{eq13}) with the replacement $V_{ts} \to V_{tb}$ cannot be used.

\section{Summary}

\hspace*{6mm} We consider the MSSM with explicit $CP$ violation in the effective two-doublet potential and the Yukawa sector of type II (MSSM II) in the framework of the scenario opposite to the decoupling limit for heavy scalars. In the case of moderate SUSY scale (a few hundreds of GeV) relatively small Higgs boson masses and large Higgs bosons mass splittings of the order of masses the $CP$ invariance of the effective potential is strongly violated. This scenario admits a relatively light charged Higgs boson $m_{H^\pm} \sim$50 GeV giving nonnegligible nonstandard contributions to the neutral mesons mixing. 

Using the low-energy approximation ($k^{2} \gg p^{2}_{i}${}) we first evaluate the $K^0$ mass splitting and the $CP$ violation parameter given by (8), (13), (14), and
(18). The yield of MSSM II amplitudes to $K^0$ mass splitting is extremely small far beyond the experimental precision and depends weakly on the possibly large ${\tt tg} \beta$=$v_2/v_1$ parameter, so the dominant contriution to $\Delta m_{LS}${} and mixing parameter $\varepsilon${} is defined by the standard $WW$-box amplitude. 
Analogously, for the $B^0_{s,d}$-mesons the contribution of MSSM II $B^0- \bar B^0$ amplitudes is insignificant but much larger in comparison with the case of $K^0$ mesons system, especially for $B^0_s$ mesons system at small charged Higgs boson mass. Charged scalar with mass 30-35 GeV is excluded if the precision of the $B^0$ mass splitting measurement equals 1\%.
Improvement of the experimental precision indicated in Table 4 by about one order of magnitude excludes the charged Higgs boson mass $m_{H^\pm}<$ 112 GeV for $B_{d}$ mesons system and $m_{H^\pm}<$ 155 GeV for $B_{s}$ mesons system. In principle the LEP2 limit could be improved by precise measurements on the $B$-factories.

Limitations on the charged scalar mass can restrict strongly the complex MSSM parameter space as illustrated by the two-dimensional contours in the variables $(\varphi, {\tt tg}\beta)$ and $(A_{t,b},
\mu)$, see Fig. 6a and 6b. Inside the contours shown in Fig. 6a which are generated for the CPX scenario ($M_{SUSY}=$500 GeV) the lightest scalar mass $m_{h_1}$ is positively defined.
At $m_{H^\pm}=$ 60 GeV only the region inside the smallest contour in the vicinity of ${\tt tg}\beta=$3 and zero phase is allowed, at the value $m_{H^\pm}=$ 70 GeV this region slightly increases, and at  $m_{H^\pm}=$ 80 GeV an additional allowed region appears at large ${\tt tg}\beta $ close to 40, which is broadening at $m_{H^\pm}=$ 90 GeV (outer contours in Fig. 6a). At large ${\tt tg}\beta \sim 45$ the lightest neutral scalar mass $m_{h_1}$ demonstrates high sensitivity to the values of $(A_{t,b}, \mu)$ parameters illustrated in Fig. 6b. The contours in Fig. 6b are generated for $m_{h_1}=$50 GeV and ${\tt tg}\beta=$45, the inner contour at the value
$m_{H^\pm}=$ 30 GeV is increasing for charged scalar mass
50, 70 and 90 GeV (outer contour). Inside the contours $m_{h_1}$ exceeds 50 GeV. The allowed regions in Fig. 6b weakly depend on the particular choice of $m_{h_1}=$50 GeV. So the light charged scalar with mass 50 --  60 GeV in the scenarios with strong mixing of $CP$-even/$CP$-odd states and large ${\tt tg}\beta$ does not leave much freedom for the choice of other complex MSSM parameters.

We do not take into account the contributions of MSSM box diagrams with quark superpartners
$q_{L,R}$, chargino $\xi^\pm$ and neutralino $\xi^0$, defined by the trilinear
MSSM Lagrangian terms $\bar q q \xi^0$ $\bar q_1 q_2 \xi^\pm$ (for example, horisontal sides of a box are $\xi^0$, and the exchange particles are
$q_{L,R}$). These Lagrangian terms include the 4$\times$4 neutralino mixing matrix elements and 2$\times$2 chargino mixing matrix elements, defined by the SUSY breaking mechanism, which are, generally speaking, complex numbers insofar as their phases cannot be removed by a redifinition of the spinor fields. The MSSM box contributions are dependent on a large parameter set involving sparticle masses and phases of matrix elements. Abundance of free parameters makes difficult the systematic estimates of MSSM contributions which can be very large or marginally small depending on a region of parameter space. This situation forces one to consider artificially simplified MSSM scenarios even for a relatively simple observables in the states $t\bar t$, $W^+ W^-$, see, for example, \cite{christova},
which are accessible for direct reconstruction on the colliders. Explicit results for MSSM contributions to the meson mixing amplitudes in the class of models with radiative generation of fermion masses can be found in \cite{diazcruz}. Following \cite{diazcruz} the MSSM contributions increase strongly at large ${\tt tg}\beta$ around 40-50, specific choice of the phases and relations between the sparticles and gaugino masses. This scenario is not overlapping with our case when MSSM box amplitudes are expected to be not large at moderate ${\tt tg}\beta$ and degenerate sparticle masses.

Our approximations are adequate for mixings at small external momenta
$p_{i}$ in comparison with the momentum $k$ in the loop. Evaluation of amplutudes at $p_{i}$ values which could be not small and probably give corrections of the order of 50\% 
to $\Delta m_{LS}${} and of the order of
$10^{-2}${} to $\varepsilon${} ( see \cite{vysocki2}) have been performed in \cite{urban},
\cite{xiao} for the general nonsupersymmetric two-doublet models with Yukawa sector of type I and II and also in \cite{dias-2} for the Yukawa sector of type III. Large loop momentum modifies Inami-Lim-Vysotsky functions \cite{urban}. The asymptotics of (21), (22) coincides with the asymptotics of modified functions in the limiting case $m_{H} \; \longrightarrow \; 0${} corresponding to the SM $WW$ box contribution only.
\begin{center}
{\bf Acknowledgements}
\end{center}
M.D. is grateful to M.Dolgopolov for useful discussion. Authors thank very much S.Bityukov for providing the code to calculate the exclusion limits. Work of M.D. was partially supported by
INTAS 03-51-4007 and NS 1685.2003.2.

\vskip 7mm

\begin{center}
\large{ \bf{Appendix}}  
\end{center}

The simplest approximation for the one-loop box diagrams in the four-fermion limit
$m_{H,W}^{2} \gg k^{2}$ ties up the two quark-boson vertices to the one four-quark interaction vertex
and the internal loop momentum $k$ is removed from boson propagators. The lack of $k$ powers in the denominators leads to a divergent amplitudes, so a cutoff is necessary at some momentum scale which is problematic to be meaningfully fixed. We are evaluating exactly the integrals over $k$ which are convergent.
The result is given by a dimensionless functions depending on the mass ratios for internal lines.
In the SM they are known as Inami-Lim-Vysotsky functions which we denote in the case of $HW$ and $HH$ amplitudes by $I_{HW-i}$ and $I_{HH-j}$ (see (\ref{eq23}), (\ref{eq29}) and
(\ref{eq22})).  

{\bf $WW$-box, see Fig.3} Details of calculation can be found in \cite{vysocki2}. The integrands for various boxes are (using the t'Hooft-Feynman gauge and omitting a constant factor)
\begin{equation}
I_{WW-1}^{p} \; = \; \frac{d^{4}k \; m_{c,t}^{4}m_{W}^{4}}{(k^{2}+m^{2}_{c,t})^{2}k^{2}
(k^{2}+m^{2}_{W})^{2}} \; \sim \; \frac{1}{k^{6}}
\end{equation}
for $cc$- and $tt$-boxes with $W$ exchanges,
\begin{equation}
I_{WW-2}^{p} \; = \; \frac{d^{4}k \; m_{c}^{2}m_{t}^{2}m_{W}^{4}}{(k^{2}+m^{2}_{c})
(k^{2}+m^{2}_{t})k^{2}(k^{2}+m^{2}_{W})^{2}} \; \sim \; \frac{1}{k^{6}}
\end{equation}
for $ct$-boxes with $W$ exchanges,
\begin{equation}
I_{WW-3}^{p} \; = \; \frac{d^{4}k \; k^{2}}{(k^{2}+m^{2}_{t})^{2}(k^{2}+m^{2}_{W})^{2}} \; \sim \;
\frac{1}{k^{2}}
\end{equation}
for $tt$-box with unphysical scalar, and
\begin{equation}
I_{WW-4}^{p} \; = \; \frac{d^{4}k}{(k^{2}+m^{2}_{t})^{2}(k^{2}+m^{2}_{W})^{2}} \; \sim \; \frac{1}
{k^{4}}
\end{equation}
for $tt$-box with $W$ boson and unphysical scalar.

The sum of these contributions gives (\ref{in-1}) and (\ref{in-2}). In the SM the four-fermion approximation for $WW$-box amplitude also converges, so keeping $W$ propagators is not a principal improvement.

{\bf $HH$-box, see Fig. 5} The integrand for $cc$ and $tt$-boxes looks as
\begin{equation}
I_{HH-1}^{p} \; = \; \frac{d^{4}k \; k^{\mu}k^{\nu} m_{c,t}^{2}}{(k^{2}+m^{2}_{c,t})^{2}
(k^{2}+m^{2}_{H})^{2}} \; \sim \; \frac{1}{k^{2}}.
\end{equation}
Exact evaluation gives the dimensionless function
\begin{equation}
I_{HH-1} \; = \; \frac{\xi_{1,2} + \xi_{6}}{(\xi_{1,2} - \xi_{6})^{2}} - \frac{2\xi_{1,2}\xi_{6}}
{(\xi_{1,2} - \xi_{6})^{3}} \ln \xi_{4,5}.
\end{equation}
For the $ct$-box the integrand has the form  
\begin{equation}
I_{HH-2}^{p} \; = \; \frac{d^{4}k \; k^{\mu}k^{\nu} m_{c}m_{t}}{(k^{2}+m^{2}_{c})(k^{2}+m^{2}_{t})
(k^{2}+m^{2}_{H})^{2}} \; \sim \; \frac{1}{k^{2}},
\end{equation}
calculation of the integral leads to  
$$
I_{HH-2} \; = \; \frac{\xi_{6}^{3}(\xi_{1} - \xi_{2}) + \xi_{6}^{2}(\xi_{2}^{2} - \xi_{1}^{2}) +
\xi_{1}^{2}\xi_{6}^{2} \ln \xi_{4} - \xi_{2}^{2}\xi_{6}^{2} \ln \xi_{5} - \xi_{1}^{2}\xi_{2}^{2}
\ln \xi_{3}}{(\xi_{6} - \xi_{1})^{2}(\xi_{6} - \xi_{2})^{2}(\xi_{1} - \xi_{2})} \; +
$$
\begin{equation}
+ \; \frac{\xi_{1}\xi_{2}\xi_{6}(\xi_{1} - \xi_{2} + 2\xi_{2} \ln \xi_{5} - 2 \xi_{1} \ln \xi_{4})}
{(\xi_{6} - \xi_{1})^{2}(\xi_{6} - \xi_{2})^{2}(\xi_{1} - \xi_{2})}.
\end{equation}
where $\xi_{1} = (\frac{m_{c}}{m_{W}})^{2}, \, \xi_{2} = (\frac{m_{t}}
{m_{W}})^{2}, \, \xi_{3} = (\frac{m_{t}}{m_{c}})^{2}, \, \xi_{4} = (\frac{m_{c}}{m_{H}})^{2}, \,
\xi_{5} = (\frac{m_{t}}{m_{H}})^{2}, \, \xi_{6} = (\frac{m_{H}}{m_{W}})^{2}$. Unlike the
SM $WW$-box the four-fermion amplitudes are divergent. 

{\bf $HW$-box, see Fig. 4} The evaluation is more or less simple in the t'Hooft-Feynman gauge
when boxes with unphysical scalars (Fig. 7) must be accounted for. The integrands are

\begin{equation}
I_{HW-1}^{p} \; = \; \frac{d^{4}k \; k^{\mu}k^{\nu} m_{c,t}^{3}}{(k^{2}+m^{2}_{c,t})^{2}
(k^{2}+m^{2}_{H})(k^{2}+m_{W}^{2})k^{2}} \; \sim \; \frac{1}{k^{4}}
\end{equation}
for  $cc$- and $tt$-boxes, and
\begin{equation}
I_{HW-2}^{p} \; = \; \frac{d^{4}k \; k^{\mu}k^{\nu} m_{c}m_{t}(m_{c} + m_{t})}{k^{2}
(k^{2}+m^{2}_{c})(k^{2}+m^{2}_{t})(k^{2}+m^{2}_{H})(k^{2} + m_{W}^{2})} \; \sim \; \frac{1}{k^{4}}
\end{equation}
for $ct$-box. Dimensionless functions can be evaluated as follows:
$$
I_{HW-1} \; = \; \frac{\xi_{1,2}^{2}\xi_{6}(1 - \ln \xi_{4,5}) + \xi_{1,2}^{2}(\ln \xi_{1,2} - 1) -
\xi_{1,2}\xi_{6}^{2} - 2 \xi_{1,2}\xi_{6} \ln \xi_{6}}{(1 - \xi_{6})(\xi_{1,2} - \xi_{6})^{2}
(\xi_{1,2} - 1)^{2}} \; +
$$
\begin{equation}
+ \; \frac{\xi_{1,2} + \xi_{6}^{2}(1 + \ln \xi_{1,2}) - \xi_{6}(1 + \ln \xi_{4,5})}{(1 - \xi_{6})
(\xi_{1,2} - \xi_{6})^{2}(\xi_{1,2} - 1)^{2}}
\end{equation}
and
$$
I_{HW-2} \; = \; \frac{\xi_{1}^{2}\xi_{2}(\ln \xi_{2} - \xi_{6} \ln \xi_{5}) - \xi_{1}^{2}\xi_{6}
\ln \xi_{6} + \xi_{1} \ln \xi_{1} (\xi_{2} + \xi_{6}) + \xi_{1}\xi_{2}^{2}\xi_{6} \ln \xi_{4}}{(1 -
\xi_{6})(\xi_{1} - \xi_{6})(\xi_{2} - 1)(\xi_{1} - 1)(\xi_{2} - \xi_{6})(\xi_{1} - \xi_{2})} \; +
$$
\begin{equation}
+ \; \frac{\xi_{1}\xi_{2}\xi_{6}^{2} \ln \xi_{3} - \xi_{1}\xi_{6}\ln \xi_{4} - \xi_{1}\xi_{2}^{2}\ln
\xi_{1} + \xi_{2}^{2}\xi_{6}\ln \xi_{6} - \xi_{2}\xi_{6}^{2} \ln \xi_{2} + \xi_{2}\xi_{6} \ln
\xi_{5}}{(1 - \xi_{6})(\xi_{1} - \xi_{6})(\xi_{2} - 1)(\xi_{1} - 1)(\xi_{2} - \xi_{6})(\xi_{1} -
\xi_{2})}.
\end{equation}
The four-fermion amplitude is divergent.
For the box diagram with unphysical scalar $G$ in Fig. 7
\begin{equation}
I_{HW-3}^{p} \; = \; \frac{d^{4}k \; k^{\mu}k^{\nu} m_{c,t}^{2}}{(k^{2}+m^{2}_{c,t})^{2}
(k^{2}+m^{2}_{H})(k^{2}+m_{W}^{2})} \; \sim \; \frac{1}{k^{2}}
\end{equation}
for $cc$- and $tt$-boxes, and
\begin{equation}
I_{HW-4}^{p} \; = \; \frac{d^{4}k \; k^{\mu}k^{\nu} m_{c}m_{t}}{(k^{2}+m^{2}_{c})(k^{2}+m^{2}_{t})
(k^{2}+m^{2}_{H})(k^{2} + m_{W}^{2})} \; \sim \; \frac{1}{k^{2}}
\end{equation}
for $ct$-boxes leading to the dimensionless functions
$$
I_{HW-3} \; = \; \frac{\xi_{1,2}^{3} - \xi_{1,2}^{3}\xi_{6} - \xi_{1,2}^{2} - \xi_{1,2}^{2} \ln
\xi_{1,2} + \xi_{1,2}^{2} \xi_{6}^{2} \ln \xi_{4,5} + \xi_{1,2}^{2} \xi_{6}^{2}}{(\xi_{6} - 1)
(\xi_{1,2} - \xi_{6})^{2}(\xi_{1,2} - 1)^{2}} \; +
$$
\begin{equation}
+ \; \frac{\xi_{1,2}\xi_{6} + 2\xi_{1,2}\xi_{6}\ln \xi_{1,2} - \xi_{1,2}\xi_{6}^{2} - 2
\xi_{1,2}\xi_{6}^{2} \ln \xi_{4,5} - \xi_{6}^{2} \ln \xi_{6}}{(\xi_{6} - 1)(\xi_{1,2} -
\xi_{6})^{2}(\xi_{1,2} - 1)^{2}}
\end{equation}
and 
$$
I_{HW-4} \; = \; \frac{\xi_{6}^{2}\ln \xi_{6}(\xi_{1} - \xi_{2}) + \xi_{1}^{2} \ln \xi_{1} (\xi_{2}
- \xi_{6}) + \xi_{1}\xi_{2}^{2}(\xi_{6}^{2} \ln \xi_{5} - \ln \xi_{2})}{(1 - \xi_{6})(\xi_{1} -
\xi_{6})(\xi_{2} - 1)(\xi_{1} - 1)(\xi_{2} - \xi_{6})(\xi_{2} - \xi_{1})} \; +
$$
\begin{equation}
+ \; \frac{\xi_{6}^{2}(\xi_{1}^{2} \ln \xi_{4} - \xi_{2}^{2} \ln \xi_{5}) + \xi_{1}^{2}\xi_{2}^{2}
\ln \xi_{3}(1 - \xi_{6}) - \xi_{2}\xi_{6}(\xi_{1}^{2}\xi_{6} \ln \xi_{4} + \xi_{2} \ln \xi_{2})}{(1
- \xi_{6})(\xi_{1} - \xi_{6})(\xi_{2} - 1)(\xi_{1} - 1)(\xi_{2} - \xi_{6})(\xi_{2} - \xi_{1})}.
\end{equation}

\vspace{2mm}

\newpage

\vspace*{40mm}

\begin{figure}[h!]
\begin{center}
\begin{picture}(10,70)
\put(-240,100){\mbox{\Large \bf $m_{H^\pm}$}}
\put(120,-55){\mbox{\Large \bf ${\tt tg} \beta$}}
\put(-200,-50){\epsfxsize=7.0cm
         \epsfysize=7.0cm \leavevmode \epsfbox{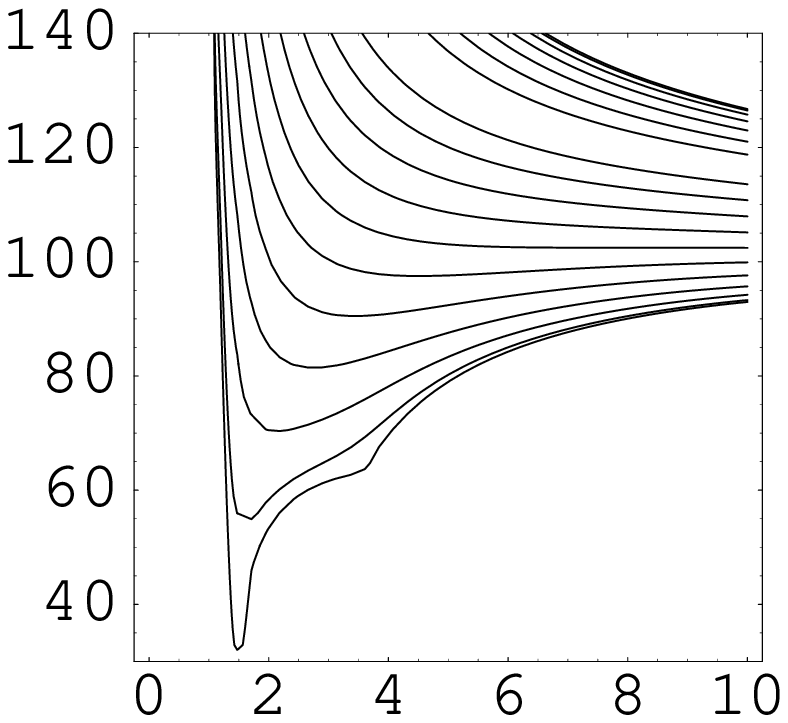}}
\put(-100,-10){\mbox{(a)}}
\put(130,-10){\mbox{(b)}}
\put(20,-50){\epsfxsize=7.0cm
         \epsfysize=7.0cm \leavevmode \epsfbox{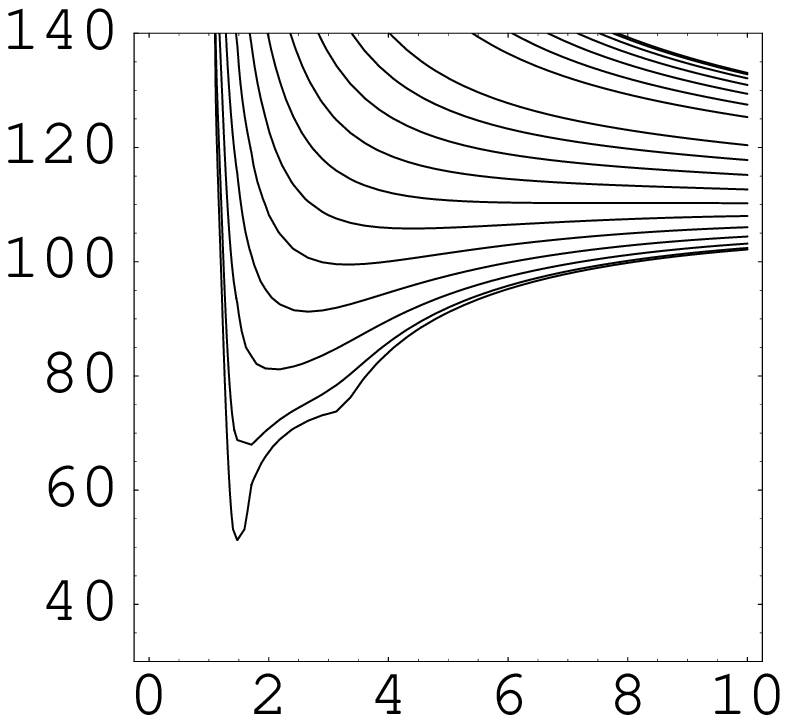}}
\end{picture}
\end{center}
\vspace{15mm}
\caption[]{\label{fg:chargedh} \small Charged Higgs boson mass in the complex MSSM II assuming that (a) the lightest neutral Higgs boson mass $m_{h_1}$ is positively defined, (b) the lightest neutral Higgs boson mass $m_{h_1}$ equals 40 GeV, as a function of ${\tt tg} \beta= v_2/v_1$. The sequence of contours is shown for the phases of
two-Higgs-doublet potential effective parameters $\lambda_6$,
$\lambda_7$ changing from zero (lower curve on both plots) to
$180^{\circ}$ (upper curve on both plots), with the step $10^{\circ}$. The CPX strong mixing scenario \cite{higgs1} is used at the scale $M_{SUSY}=$500 GeV. Below the contour for a definite phase value the lightest neutral Higgs boson mass (a) is not positively defined (b) is less than 40 GeV.}
\end{figure}

\newpage

\vspace*{40mm}

\begin{figure}[h!]
\begin{center}
\begin{picture}(10,70)
\put(-240,100){\mbox{\Large \bf $m_{H^\pm}$}}
\put(120,-55){\mbox{\Large \bf ${\tt tg} \beta$}}
\put(-200,-50){\epsfxsize=7.0cm
         \epsfysize=7.0cm \leavevmode \epsfbox{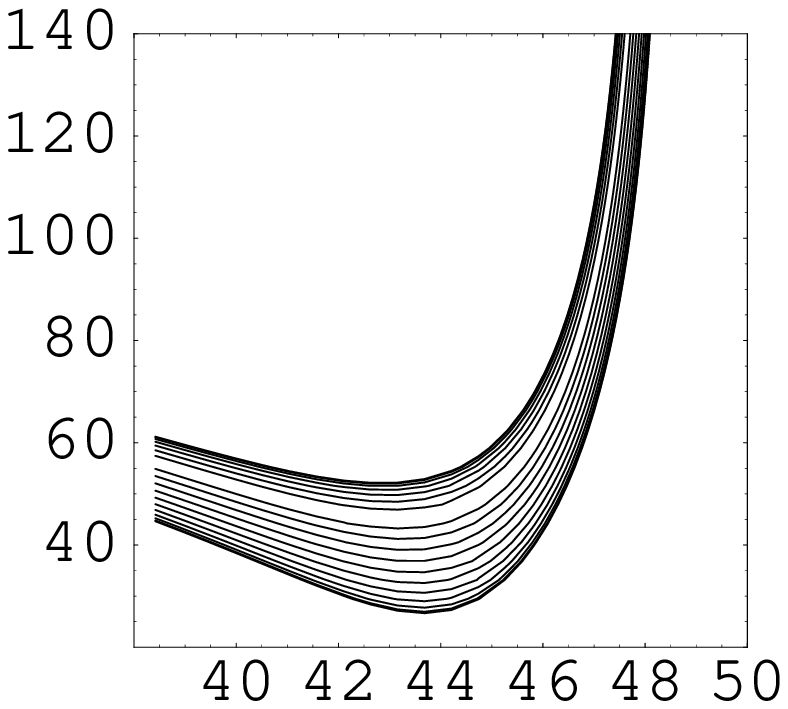}}
\put(-50,-10){\mbox{(a)}}
\put(130,-10){\mbox{(b)}}
\put(20,-50){\epsfxsize=7.0cm
         \epsfysize=7.0cm \leavevmode \epsfbox{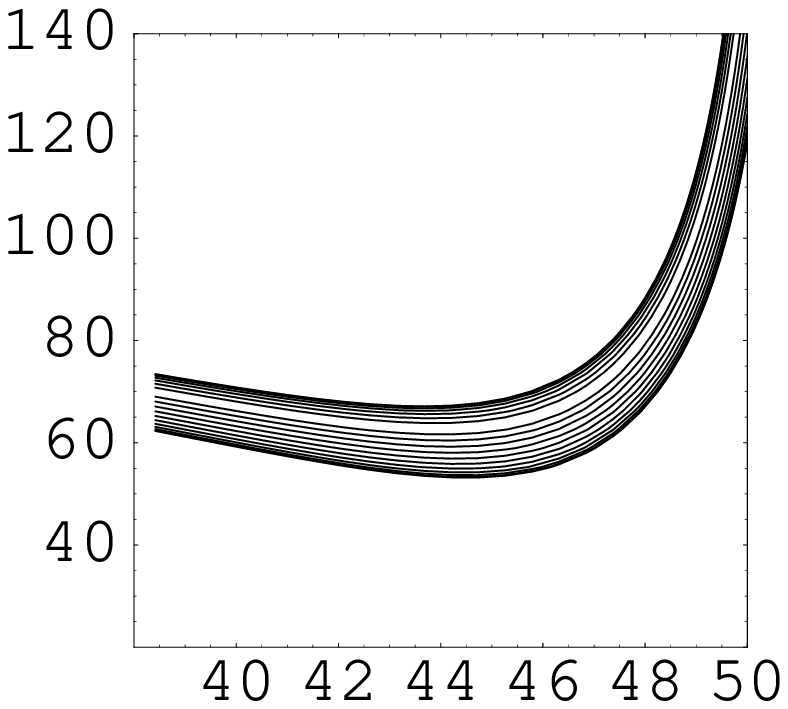}}
\end{picture}
\end{center}
\vspace{15mm}
\caption[]{\label{fg:chargedh2} \small Charged Higgs boson mass in the complex MSSM II assuming that the lightest neutral Higgs boson mass $m_{h_1} \sim$50 GeV at large values of ${\tt tg} \beta$ parameter. The curves are insignificantly sensitive to the limit imposed on $m_{h_1}$. The sequence of contours is shown for the phases of two-Higgs-doublet potential effective parameters $\lambda_6$,
$\lambda_7$ changing from zero (lower curve on both plots) to $180^{\circ}$ (upper curve on both plots), with the step $10^{\circ}$. The strong mixing scenario close to CPX \cite{higgs1} is used at the scale $M_{SUSY}=$500 GeV (a) $A_{t,b}=$890 GeV, $\mu= $2000 GeV, (b) $A_{t,b}=$890 GeV, $\mu= $1900 GeV}
\end{figure}

\newpage

\begin{figure}[h!]
\begin{center}
\vspace{10mm}
\begin{picture}(10,10)
\put(-105,-80){\epsfxsize=7.0cm
         \epsfysize=4.8cm \leavevmode \epsfbox{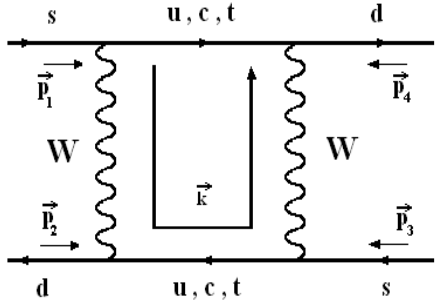}}
\end{picture}
\end{center}
\vspace {25mm}
\caption[]{\label{fig3} The GIM mechanism for $K^{0} \longrightarrow {\tilde K}^{0}${} mixing.}
\end{figure}

\vspace{25mm}

\begin{figure}[h!]
\begin{center}
\begin{picture}(10,10)
\put(-105,-50){\epsfxsize=7.0cm
         \epsfysize=4.8cm \leavevmode \epsfbox{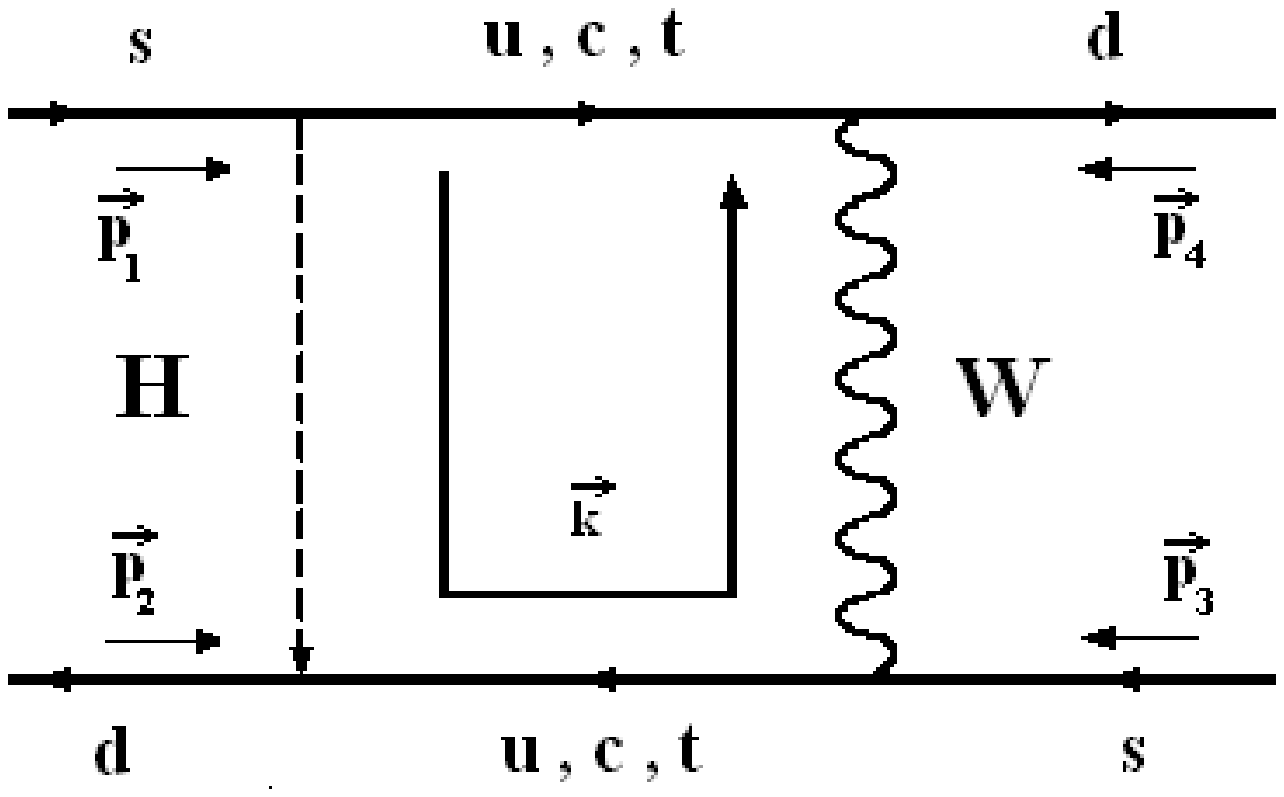}}
\end{picture}
\end{center}
\vspace{10mm}
\caption[]{\label{fig4} The $HW$-box diagram for $K^{0} \longrightarrow {\tilde K^{0}}${} mixing.}
\end{figure}

\vspace{29mm}

\begin{figure}[h!]
\begin{center}
\begin{picture}(10,10)
\put(-105,-50){\epsfxsize=7.0cm
         \epsfysize=4.8cm \leavevmode \epsfbox{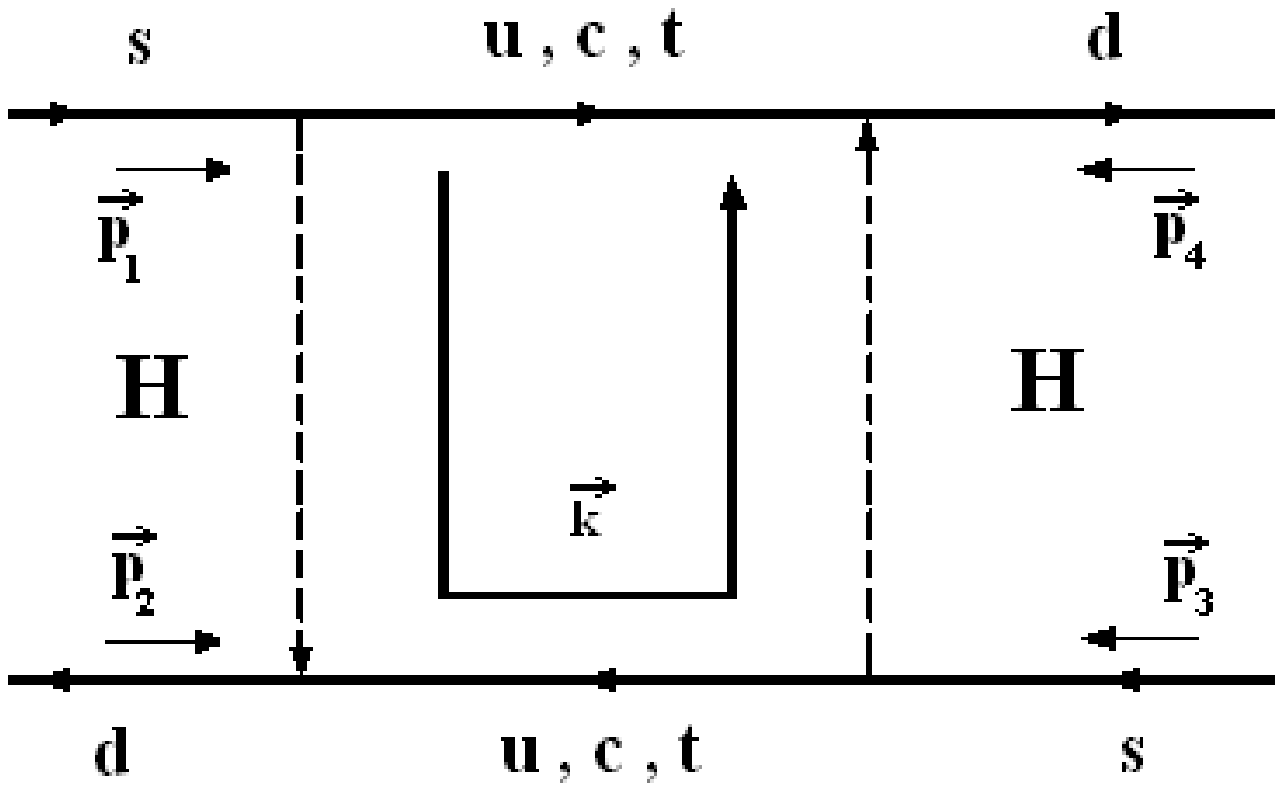}}
\end{picture}
\end{center}
\vspace{10mm}
\caption[]{\label{fig5} The $HH$-box diagram for $K^{0} \longrightarrow {\tilde K^{0}}${} mixing.}
\end{figure}

\newpage

\vspace*{40mm}

\begin{figure}[h!]
\begin{center}
\begin{picture}(10,70)
\put(-210,120){\mbox{\Large \bf $\varphi$}}
\put(-40,-50){\mbox{\Large \bf ${\tt tg} \beta$}}
\put(160,-55){\mbox{\Large \bf $A_{t,b}$}}
\put(10,120){\mbox{\Large \bf $\mu$}}
\put(-200,-50){\epsfxsize=7.0cm
         \epsfysize=7.0cm \leavevmode \epsfbox{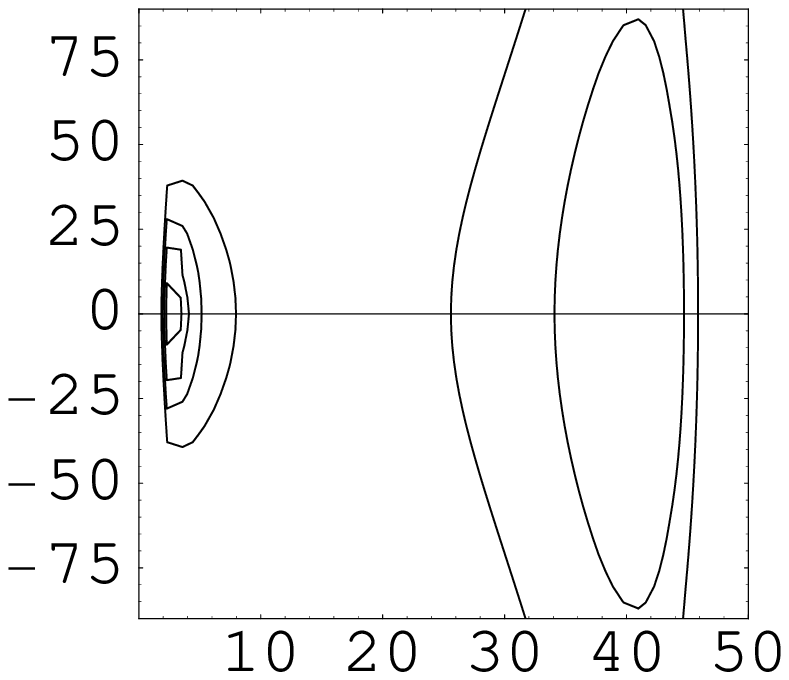}}
\put(-100,-10){\mbox{(a)}}
\put(130,-10){\mbox{(b)}}
\put(20,-50){\epsfxsize=7.0cm
         \epsfysize=7.0cm \leavevmode \epsfbox{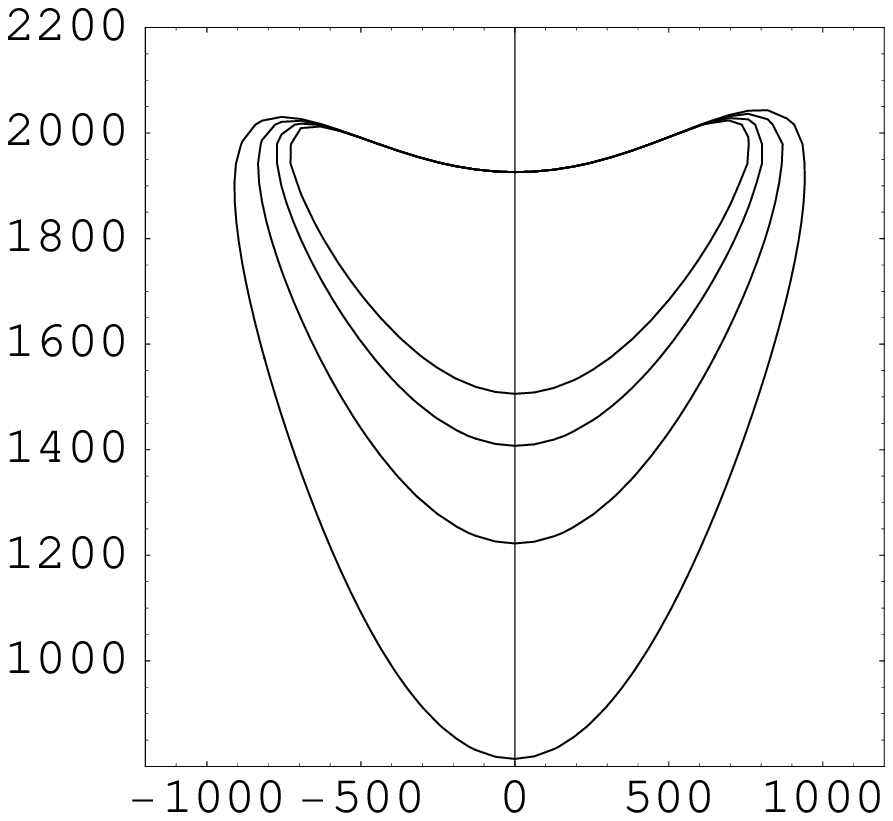}}
\end{picture}
\end{center}
\vspace{15mm}
\caption[]{\label{twodim} (a) Regions of lightest Higgs boson mass $m_{h_1}$ positively defined, (b) Regions of $m_{h_1}>$50 GeV, see the text for details.}
\end{figure}

\vspace{39mm}

\begin{figure}[h!]
\begin{center}
\begin{picture}(10,10)
\put(-105,-50){\epsfxsize=8.0cm
         \epsfysize=6cm \leavevmode \epsfbox{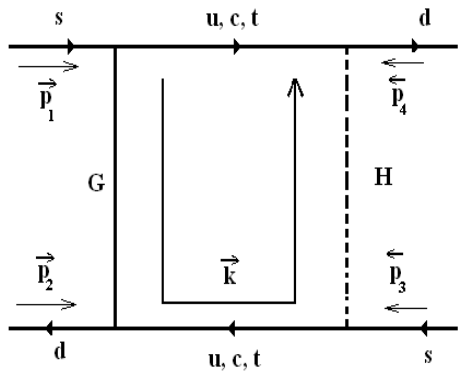}}
\end{picture}
\end{center}
\vspace{10mm}
\caption[]{\label{fig7} The box diagram with unphysical scalar mode for $K^{0} \longrightarrow {\tilde K^{0}}${} mixing.}
\end{figure}

\newpage

\begin{table}[h!]
\caption{\small $K^0$ mesons mass spitting in the SM and MSSM II. The contributions of amplitudes shown in Figs.3,4 and 5 to $\Delta M_{LS}${} are indicated in the second, third and fourth lines.}
\begin{center}
\begin{tabular}[c]{||c||c|c|c|c|c|c|c||} \hline
$m_{H^\pm}$, GeV & 50 & 100 & 150 & 200 & 250 & 300 & 500 \\ \hline \hline
$\Delta m_{LS}^{WW} \times 10^{15}$, ГэВ & \multicolumn{7}{|c||}{2.721 ($\Delta m_{LS}^{exp} =
(3.482 \pm 0.013) \times 10^{-15}$, ГэВ)} \\ \hline
$\Delta m_{LS}^{HW} \times 10^{19}$, ГэВ & -14.22 & -4.44 & -2.19 & -1.31 & -0.87 & -0.63 & -0.24 \\
\hline
$\Delta m_{LS}^{HH} \times 10^{22}$, ГэВ & 61.50 & 15.79 & 7.15 & 4.07 & 2.63 & 1.84 & 0.67 \\
\hline 
\end{tabular}
\end{center}

\end{table}


\begin{table}[h!]
\caption{Indirect $CP$ violation for $K^0$ mesons in the SM (second line) and MSSM II (third line).}
\begin{center}
\begin{tabular}[c]{||c||c|c|c|c|c|c|c||} \hline
$m_{H}$, GeV & 50 & 100 & 150 & 200 & 250 & 300 & 500 \\ \hline \hline
$|\varepsilon|_{WW} \times 10^{3}$ & \multicolumn{7}{|c||}{2.0523 ($\varepsilon_{LS}^{exp} = (2.232
\pm 0.007) \times 10^{-3}$)} \\ \hline
$|\varepsilon|_{tot} \times 10^{3}$ & 2.0419 & 2.0472 & 2.0493 & 2.0503 & 2.0509 & 2.0513 & 2.0519
\\ \hline 
\end{tabular}
\end{center}

\end{table}


\begin{table}[h!]
\caption{Indirect $CP$ violation and mass splitting for $K^0$ mesons in the SM at different values of $\eta_{1}$. }
\begin{center}
\begin{tabular}[c]{||c||c|c|c|c|c||} \hline
$\eta_{1}$ & 1.1 & 1.2 & 1.3 & 1.4 & 1.5 \\ \hline \hline
$\Delta m_{LS}^{exp} \times 10^{15}$, GeV$^{-2}$ & \multicolumn{5}{|c||}{$3.48 \pm 0.01$} \\ \hline
$\varepsilon_{LS}^{exp} \times 10^{3}$ & \multicolumn{5}{|c||}{$2.23 \pm 0.01$} \\ \hline
$\Delta m_{LS}^{SM} \times 10^{15}$, GeV$^{-2}$ & 2.31 & 2.51 & 2.72 & 2.93 & 3.14\\ \hline
$\varepsilon_{LS}^{SM} \times 10^{3}$ & 2.48 & 2.25 & 2.10 & 1.88 & 1.73 \\ \hline
\end{tabular}
\end{center}

\end{table}

\newpage

\begin{table}[h!]
\caption{$B^0_{d}$- and $B^0_{s}$-mesons mass splitting in the SM and MSSM II.
Lines from two to seven indicate various contributions to $\Delta M_{LS}^{B_{d, s}}${}.}
\begin{center}
\begin{tabular}[c]{||c||c|c|c|c|c|c|c||} \hline
$m_{H}$, GeV & 50 & 100 & 150 & 200 & 250 & 300 & 500 \\ \hline \hline
$\Delta m_{B_{d}}^{WW} \times 10^{13}$, GeV & \multicolumn{7}{|c||}{2.11 ($\Delta m_{B_{d}}^{exp} =
(3.33 \pm 0.03) \times 10^{-13}$, ГэВ)} \\ \hline
$\Delta m_{B_{d}}^{HW} \times 10^{16}$, GeV & -17.05 & -11.99 & -8.70 & -6.57 & -5.14 & -4.13 &
-2.09 \\ \hline
$\Delta m_{B_{d}}^{HH} \times 10^{17}$, GeV & 3.97 & 2.87 & 2.11 & 1.59 & 1.24 & 0.99 & 0.47 \\
\hline 
$\Delta m_{B_{s}}^{WW} \times 10^{12}$, GeV & \multicolumn{7}{|c||}{9.3 ($\Delta m_{B_{s}}^{exp} =
11.4^{+0.2}_{-0.1} \times 10^{-12}$, GeV)} \\ \hline
$\Delta m_{B_{s}}^{HW} \times 10^{14}$, GeV & -8.21 & -5.67 & -4.08 & -3.07 & -2.39 & -1.92 & -0.97
\\ \hline
$\Delta m_{B_{s}}^{HH} \times 10^{15}$, GeV & 1.83 & 1.31 & 0.96 & 0.72 & 0.56 & 0.45 & 0.21 \\
\hline
\end{tabular}
\end{center}

\end{table}


\begin{table}[h!]
\caption{Relative contribtions to the mass splitting provided by $HW$-boxes and SM $WW$-boxes. $R_{B_{s,d}}$ is the ratio of $|\Delta m_{B_{d, s}}^{HW} / \Delta
m_{B_{d, s}}^{WW}|$.}
\begin{center}
\begin{tabular}[c]{||c||c|c|c|c|c|c|c||} \hline
$m_{H}$, GeV & 50 & 75 & 100 & 125 & 150 & 175 & 200 \\ \hline \hline
$\Delta m_{B_{d}}^{WW} \times 10^{13}$, GeV & \multicolumn{7}{|c||}{2.11 ($\Delta m_{B_{d}}^{exp} =
(3.33 \pm 0.03) \times 10^{-13}$, ГэВ)} \\ \hline
$\Delta m_{B_{d}}^{HW} \times 10^{16}$, GeV & -17.05 & -14.28 & -11.99 & -10.16 & -8.70 & -7.57 &
-6.57 \\ \hline
$R_{B_{d}}$  & 0.0081 & 0.0068 & 0.0057 & 0.0048 & 0.0041 & 0.0036 & 0.0031 \\ \hline 
$\Delta m_{B_{s}}^{WW} \times 10^{12}$, GeV & \multicolumn{7}{|c||}{9.3 ($\Delta m_{B_{s}}^{exp} =
11.4^{+0.2}_{-0.1} \times 10^{-12}$, GeV)} \\ \hline
$\Delta m_{B_{s}}^{HW} \times 10^{14}$, GeV & -8.21 & -6.79 & -5.67 & -4.78 & -4.08 & -3.54 & -3.07
\\ \hline
$R_{B_{s}}$ & 0.0088 & 0.0073 & 0.0061 & 0.0051 & 0.0044 & 0.0038 & 0.0033 \\ \hline
\end{tabular}
\end{center}

\end{table}

\end{document}